\documentclass[twocolumn,showpacs,preprintnumbers,amsmath,amssymb,superscriptaddress]{revtex4}

\usepackage{graphicx}
\usepackage{epstopdf}
\usepackage{dcolumn}
\usepackage{bm}

\begin{document}
　
\title{$^{23}$Na nuclear spin-lattice relaxation studies of Na$_2$Ni$_2$TeO$_6$}

\author{Y. Itoh}
 \affiliation{Department of Physics, Graduate School of Science, Kyoto Sangyo 
University, Kamigamo-Motoyama, Kita-ku Kyoto 603-8555, Japan} 　

\date{\today}

\begin{abstract}
We report on $^{23}$Na NMR studies of a honeycomb lattice antiferromagnet Na$_2$Ni$_2$TeO$_6$ by $^{23}$Na nuclear spin-echo techniques. 
The $^{23}$Na nuclear spin-lattice relaxation rate 1/$^{23}T_1$ exhibits critical divergence near a N\'{e}el temperature $T_\mathrm{N}$ = 26 K, a narrow critical region, and a critical exponent $w$ = 0.34 in 1/$^{23}T_1$ $\propto$ ($T/T_\mathrm{N}$ - 1)$^{-w}$ for Na$_2$Ni$_2$TeO$_6$, and $T_\mathrm{N}$ = 18 K for Na$_2$(Ni$_{0.5}$Cu$_{0.5}$)$_2$TeO$_6$.  
Although the uniform magnetic susceptibility of Na$_2$Ni$_2$TeO$_6$ exhibits a broad maximum at 35 K characteristic of low dimensional spin systems, the NMR results indicate three dimensional critical phenomenon around the N\'{e}el temperature.        
\end{abstract}

\maketitle
 
\section{Introduction} 
Na$_2$Ni$_2$TeO$_6$ is a quasi-two dimensional honeycomb lattice antiferromagnet~\cite{Miura,SL,Ch}. The crystal structure of Na$_2$Ni$_2$TeO$_6$ consists of the stacking of Na and (Ni/Te)O$_6$ layers ($P6_3/mcm$)~\cite{SL,Ch}. The N\'{e}el temperature of $T_\mathrm{N}$ $\approx$ 27 K was estimated from measurements of specific heat and the derivative of uniform magnetic susceptibility~\cite{Ch}. 
The magnetic susceptibility takes a broad maximum around 34 K~\cite{SL,Ch}.  
The Weiss temperature of $\theta$ = - 32 K and the superexchange interaction of $J/k_\mathrm{B}$ =  - 45 K were estimated from the analysis of Curie-Weiss law fit and a high temperature series expansion~\cite{Ch}.   
Although the Ni$^{2+}$ ion must carry a local moment of $S$ = 1 on the honeycomb lattice, the large effective moment $\mu_\mathrm{eff}$ = 3.446$\mu_\mathrm{B}$ could not be explained by spin $S$ = 1 with a $g$-factor of $g$ = 2 ~\cite{Ch}. The $g$-factor must be larger than 2~\cite{SL}, or a Ni$^{3+}$ ion and the intermediate state might be realized because of the tunable valence of Te$^{4+}$ and Te$^{6+}$~\cite{Ch}.

Spin frustration effects on a honeycomb lattice have renewed our interests, since the discovery of a possible spin liquid state in a spin-3/2 antiferromagnet~\cite{Azuma}. 
Various magnetic ground states are competitive with each other on the honeycomb lattice~\cite{J1J2J3}. 

In this paper, we report on $^{23}$Na NMR studies of Na$_2$Ni$_2$TeO$_6$ and Na$_2$(Ni$_{0.5}$Cu$_{0.5}$)$_2$TeO$_6$ polycrystalline samples. 
Na$_2$(Ni$_{0.5}$Cu$_{0.5}$)$_2$TeO$_6$ still belongs to the same space group $P6_3/mcm$ as Na$_2$Ni$_2$TeO$_6$~\cite{SL,Mor}. 
For the Cu substitution, we expected a possible enhancement of quantum effects from $S$ = 1 to 1/2.
Since the solubility limit in the honeycomb lattice Na$_2$(Ni$_{1-x}$Cu$_{x}$)$_2$TeO$_6$ is about $x$ = 0.6~\cite{Mor}, 
we selected the half Cu-substituted sample being away from the phase boundary.  
We observed three dimensional critical phenomenon in the $^{23}$Na nuclear spin-lattice relaxation rate 1/$^{23}T_1$ near $T_\mathrm{N}$ = 26 K for Na$_2$Ni$_2$TeO$_6$ and $T_\mathrm{N}$ = 18 K for Na$_2$(Ni$_{0.5}$Cu$_{0.5}$)$_2$TeO$_6$. 
The broad maximum of uniform magnetic susceptibility is not the onset of magnetic long range ordering. 
In the antiferromagnetic state of Na$_2$Ni$_2$TeO$_6$, we observed 1/$^{23}T_1$$\propto$$T^3$ which indicates conventional spin-wave scattering.  
 
 \section{Experiments}

\begin{figure}[b]
 \begin{center}
 \includegraphics[width=0.75\linewidth]{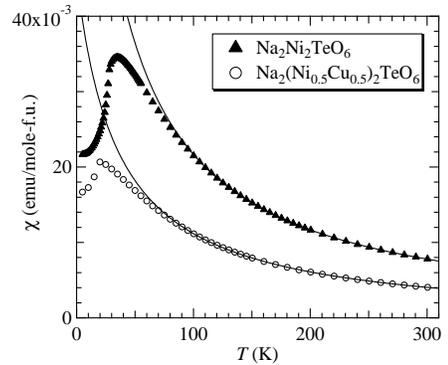}
 \end{center}
 \caption{\label{XT}
Uniform magnetic susceptibility $\chi$ of Na$_2$Ni$_2$TeO$_6$ and Na$_2$(Ni$_{0.5}$Cu$_{0.5}$)$_2$TeO$_6$.   
Solid curves are the results from least squares fits by a Curie-Weiss law.
 }
 \end{figure} 
 
Powder samples of Na$_2$Ni$_2$TeO$_6$ have been synthesized by a conventional solid-state reaction method. 
Appropriate amounts of NiO, TeO$_6$ and Na$_2$CO$_3$ were mixed, palletized and fired a few times at 800 - 860$^\circ$C and finally at 900$^\circ$C for 24 hours in air. 
The products were confirmed to be in a single phase from measurements of powder X-ray diffraction patterns. 
Magnetic susceptibility $\chi$ at 1.0 T was measured by a superconducting quantum interference device (SQUID) magnetometer. 
Powder samples of Na$_2$(Ni$_{0.5}$Cu$_{0.5}$)$_2$TeO$_6$ were previously synthesized and characterized~\cite{Mor}. 

A phase-coherent-type pulsed spectrometer was utilized to perform the $^{23}$Na NMR (nuclear spin $I$ = 3/2) experiments in an external magnetic field of 7.4847 T.    
The NMR frequency spectra were obtained from Fourier transformation of the $^{23}$Na nuclear spin-echoes.   
The $^{23}$Na nuclear spin-lattice relaxation curves $^{23}p(t) = 1-E(t)/E(\infty)$ (recovery curves) were obtained by an inversion recovery technique as a function of time $t$ after an inversion pulse,  where the nuclear spin-echoes $E(t)$, $E(\infty)[\equiv E(10T_1)]$ and $t$ were recorded.   

\section{Experimental results and discussions}

\subsection{Uniform magnetic susceptibility}
 
Figure~\ref{XT} shows uniform magnetic susceptibility $\chi$ of Na$_2$Ni$_2$TeO$_6$ and Na$_2$(Ni$_{0.5}$Cu$_{0.5}$)$_2$TeO$_6$.
The solid curves are the results from least squares fits by a Curie-Weiss law.
We estimated the Weiss temperature $\theta$ = - 27 K and the effective moment $\mu_\mathrm{eff}$ = 3.4$\mu_\mathrm{B}$ for Na$_2$Ni$_2$TeO$_6$, which agree with the previous report~\cite{Ch}, and $\theta$ = - 35 K and $\mu_\mathrm{eff}$ = 2.5$\mu_\mathrm{B}$ for Na$_2$(Ni$_{0.5}$Cu$_{0.5}$)$_2$TeO$_6$. 
If the $g$-factor is $g$ = 2, then $S$ = 1 and $S$ = 1/2 lead to $\mu_\mathrm{eff}$ = 2.83$\mu_\mathrm{B}$ and 1.73$\mu_\mathrm{B}$, respectively. 
$\chi$ deviates below about 100 K from the Curie-Weiss law and takes a broad maximum at 35 K in Na$_2$Ni$_2$TeO$_6$.  
$\chi$ drops below about 20 K in Na$_2$(Ni$_{0.5}$Cu$_{0.5}$)$_2$TeO$_6$.  

\subsection{NMR spectrum and recovery curves}

\begin{figure}[h]
 \begin{center} 
\includegraphics[width=0.85\linewidth]{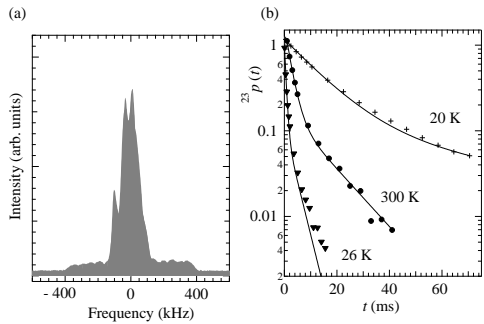}
 \end{center}
 \caption{\label{NMR}
(a) Fourier-transformed $^{23}$Na NMR spectrum at 84.670 MHz and at 300 K.  
(b) $^{23}$Na nuclear spin-lattice relaxation curves $^{23}p(t)$ at a central frequency. Solid curves are the results from least squares fits by eq.~(\ref{eq1}). 
 }
\end{figure} 

Figure~\ref{NMR}(a) shows the Fourier-transformed spectrum of $^{23}$Na spin-echoes at a Larmor frequency of 84.670 MHz and at 300 K. 
The central transition line of $Iz$ = 1/2 $\leftrightarrow$ -1/2 is affected by a nuclear quadrupole interaction~\cite{Abragam}. 
The linewidth is about 150 kHz.
 The precise value of the Knight shift could not be determined within the present studies.  

Figure~\ref{NMR}(b) shows the recovery curves $^{23}p(t)$ with varying temperature. The solid curves are the results from least-squares fits by a theoretical multi-exponential function for a central transition line ($I_z$ = 1/2 $\leftrightarrow$ -1/2)  
\begin{equation} 
^{23}p(t)=p(0)\{0.1e^{-t/^{23}T_{\rm 1}}+0.9e^{-6t/^{23}T_{\rm 1}}\},
\label{eq1}
\end{equation}
where $p(0)$ and a $^{23}$Na nuclear spin-lattice relaxation time $^{23}T_1$ are fit parameters.  
The theoretical function of eq.~(\ref{eq1}) well reproduces the experimental recovery data. 
Thus, the assignment of the exciting spectrum to the central transition line is justified $a$ $posteriori$, too. 

\subsection{Na$_2$Ni$_2$TeO$_6$}

\begin{figure}[h]
 \begin{center}
 \includegraphics[width=0.80\linewidth]{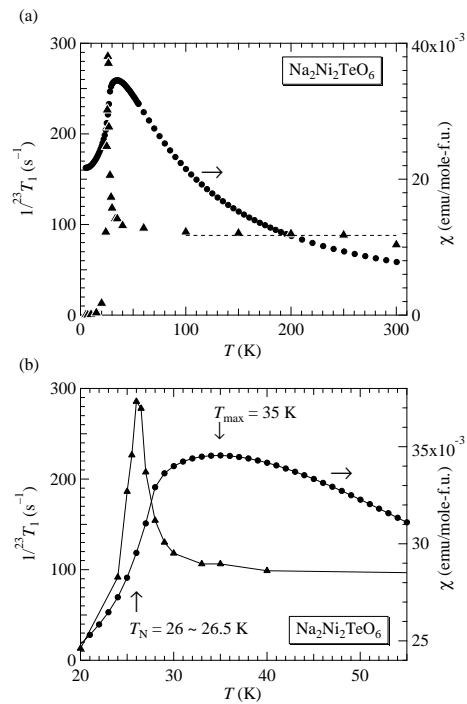}
 \end{center}
 \caption{\label{f3}
(a) 1/$^{23}T_1$ and uniform magnetic susceptibility $\chi$ against temperature. 
1/$^{23}T_1$ shows a critical divergence near $T_\mathrm{N}$ = 26 $\sim$ 26.5 K and levels off above about 100 K.   
The broken line indicates 1/$^{23}T_{1\infty}$ = 88 s$^{-1}$. 
(b) 1/$^{23}T_1$ and $\chi$ against temperature in enlarged scales. Solid curves are visual guides. 
}
 \end{figure}     
 
\begin{figure}[h]
 \begin{center}
 \includegraphics[width=0.80\linewidth]{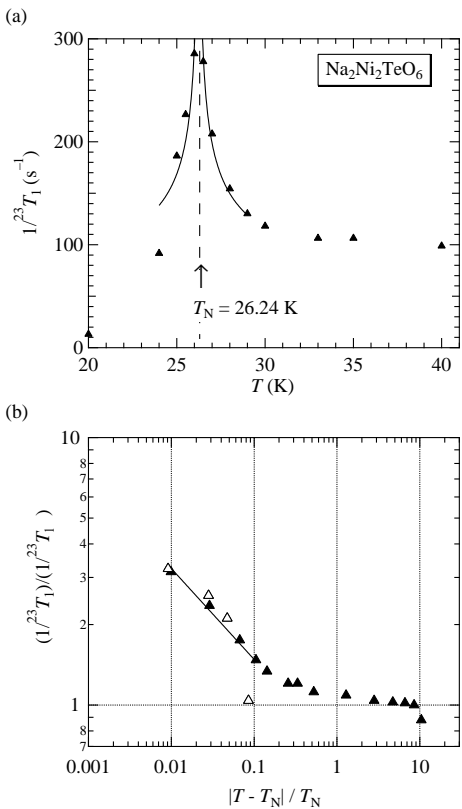}
 \end{center}
 \caption{\label{f4}
(a) 1/$^{23}T_1$ against temperature. The solid curve is the result from a least squares fit by eq.~(\ref{T1cr}). 
The N\'{e}el temperature and the critical exponent were estimated to be $T_\mathrm{N}$ = 26.24 K and $w$ = 0.34, respectively.   
(b) Log-log plots of normalized (1/$^{23}T_1$)/(1/$^{23}T_{1\infty}$) against reduced temperature $\left| T - T_\mathrm{N}\right|$/$T_\mathrm{N}$. Closed and open triangles are 1/$^{23}T_1$ above and below $T_\mathrm{N}$, respectively.  
 The solid line indicates the result from a least squares fit by eq.~(\ref{T1cr}). 
 }
 \end{figure}  
     
\begin{figure}[h]
 \begin{center}
 \includegraphics[width=0.80\linewidth]{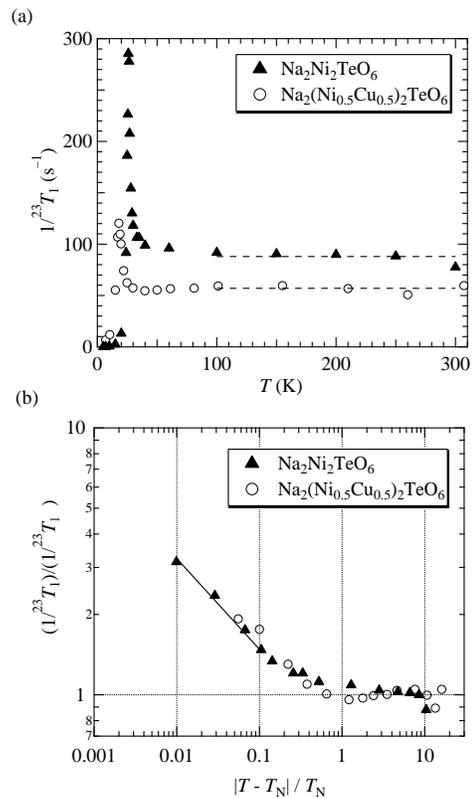}
 \end{center}
 \caption{\label{f5}
(a) 1/$^{23}T_1$ against temperature for Na$_2$Ni$_2$TeO$_6$ and Na$_2$(Ni$_{0.5}$Cu$_{0.5}$)$_2$TeO$_6$.  
The broken lines indicate 1/$^{23}T_{1\infty}$ = 88 and 57 s$^{-1}$.
(b) Log-log plots of normalized (1/$^{23}T_1$)/(1/$^{23}T_{1\infty}$) against reduced temperature $\left| T - T_\mathrm{N}\right|$/$T_\mathrm{N}$ for Na$_2$Ni$_2$TeO$_6$ ($T_\mathrm{N}$ = 26.24 K) and Na$_2$(Ni$_{0.5}$Cu$_{0.5}$)$_2$TeO$_6$  ($T_\mathrm{N}$ = 18 K). 
 The solid line is eq.~(\ref{T1cr}) with the critical exponent of $w$ = 0.34.    
 }
 \end{figure}
 
Figures~\ref{f3}(a) and (b) show 1/$^{23}T_1$ and uniform magnetic susceptibility $\chi$ against temperature.  
1/$^{23}T_1$ takes 1/$^{23}T_{1\infty}$ = 88 s$^{-1}$ above about 100 K and 
shows a divergence at 26 $\sim$ 26.5 K which can be assigned to the N\'{e}el temperature $T_\mathrm{N}$.
Thus, the broad maximum of the magnetic susceptibility $\chi$ at 35 K is not due to the antiferromagnetic long range ordering but due to a low dimensional short range correlation developing on the honeycomb lattice antiferromagnets~\cite{Motome}. 
The result is consistent with the specific heat measurements~\cite{Ch}. 

Figure~\ref{f4}(a) shows 1/$^{23}T_1$ against temperature and the result (the solid curve) from a least-squares fit by  
\begin{eqnarray}   
{1\over {^{23}T_{1}}} = {C\over {^{23}T_{1\infty}}}{1\over {|T/T_\mathrm{N} - 1|}^{w}},
\label{T1cr}
\end{eqnarray}
where a constant $C$, a N\'{e}el temperature $T_\mathrm{N}$, and a critical exponent $w$ are fit parameters. The fitting results were $T_\mathrm{N}$ = 26.24 K and $w$ = 0.34.  

A mean field theory for a three dimensional isotropic Heisenberg antiferromagnet gives $w$ = 1/2~\cite{Moriya}. A dynamic scaling theory gives $w$ = 1/3 for a three dimensional isotropic Heisenberg model~\cite{HH} and $w$ = 2/3 for a three dimensional uniaxial anisotropic Heisenberg model~\cite{RW}. The exponent of $w$ = 0.34 indicates that Na$_2$Ni$_2$TeO$_6$ in the critical region is described by a three dimensional dynamical spin susceptibility. 
In passing, CuO exhibits a similar $w$ = 0.33, a broad maximum in $\chi$ at 540 K, and $T_\mathrm{N}$ = 230 K~\cite{CuO}.  

Figure~\ref{f4}(b) shows log-log plots of normalized (1/$^{23}T_1$)/(1/$^{23}T_{1\infty}$) against reduced temperature $\left| T - T_\mathrm{N}\right|$/$T_\mathrm{N}$. The solid line indicates the result from a least squares fit by eq.~(\ref{T1cr}). 

The onset of increase in the NMR relaxation rate near $T_\mathrm{N}$ empirically categorizes critical regions. 
The region of $\left| T - T_\mathrm{N}\right|$/$T_\mathrm{N}\leq$ 10 has been assigned to the renormalized classical regime with a divergent magnetic correlation length toward $T$ = 0 K~\cite{Itoh}.   
The region of $\left| T - T_\mathrm{N}\right|$/$T_\mathrm{N}\leq$ 1.0 has been assigned to the three dimensional critical regime with a divergent magnetic correlation length toward $T_\mathrm{N}$. 
Thus, the narrow critical region of $\left| T - T_\mathrm{N}\right|$/$T_\mathrm{N}\leq$ 1 also empirically categorizes Na$_2$Ni$_2$TeO$_6$ to the three dimensional critical regime. 
 
 At high temperatures of $T \gg J$, the spin system is in the exchange narrowing limit. Then, 1/$^{23}T_1$ is expressed by 
\begin{eqnarray}   
    {1\over {^{23}T_{1\infty}}} = \sqrt{2\pi}{{S(S+1)}\over {3}}{{z_n\bigl(^{23}\gamma_n A\bigr)^2}\over {\omega_{ex}}},
    \label{T1inf}
    \label{wex}
\end{eqnarray}
    \begin{eqnarray}    
    {\omega_{ex}}^2 = {2\over {3}}S(S+1)z\Bigl({J\over {\hbar}}\Bigr)^2,
    \label{wex}
\end{eqnarray}
where $^{23}\gamma_n/2\pi$ = 11.262 MHz/T is the $^{23}$Na nuclear gyromagnetic ratio, $A$ is a hyperfine coupling constant,
and $\omega_{ex}$ is an exchange frequency~\cite{Moriya2}. 
$z_n$ is the number of Ni ions nearby a $^{23}$Na nuclear. $z$ is the number of the nearest neighbor Ni ions. 
Assuming $J$ = 45 K,~\cite{Ch} $S$ = 1, and $z$ = 3, we obtained $\omega_{ex}$ = 12 $\times$ 10$^{12}$ s$^{-1}$. 
From eq.~(\ref{T1inf}) with 1/$^{23}T_{1\infty}$  = 88 s$^{-1}$, we derived the hyperfine coupling constant $A$ = 2.0 kOe/$\mu_\mathrm{B}$,
which is nearly the same as that of Na$_3$Cu$_2$SbO$_6$~\cite{Lue}.
 
\subsection{Na$_2$(Ni$_{0.5}$Cu$_{0.5}$)$_2$TeO$_6$}
  
Figure~\ref{f5}(a) shows 1/$^{23}T_1$ against temperature for Na$_2$Ni$_2$TeO$_6$ and Na$_2$(Ni$_{0.5}$Cu$_{0.5}$)$_2$TeO$_6$. 
For the half substitution of Cu for Ni, 1/$^{23}T_{1\infty}$ and $T_\mathrm{N}$ decrease to 57 s$^{-1}$ and 18 K, respectively. 
Extrapolating linearly $T_\mathrm{N}$ with $\Delta T_\mathrm{N}$ = - 8 K per half Cu to full Cu substitution, 
one may infer $T_\mathrm{N}$ = 10 K of a hypothetical spin-1/2 honeycomb lattice “Na$_2$Cu$_2$TeO$_6$," 
although actual Na$_2$Cu$_2$TeO$_6$ is known to be monoclinic and an alternating spin chain system~\cite{Xu,M1}.

Figure~\ref{f5}(b) shows log-log plots of normalized (1/$^{23}T_1$)/(1/$^{23}T_{1\infty}$) against reduced temperature $\left| T - T_\mathrm{N}\right|$/$T_\mathrm{N}$ for Na$_2$Ni$_2$TeO$_6$ ($T_\mathrm{N}$ = 26.24 K) and Na$_2$(Ni$_{0.5}$Cu$_{0.5}$)$_2$TeO$_6$  ($T_\mathrm{N}$ = 18 K). 
The solid line indicates eq.~(\ref{T1cr}) with the critical exponent of $w$ = 0.34. 
The critical region of Na$_2$(Ni$_{0.5}$Cu$_{0.5}$)$_2$TeO$_6$ is still narrow as the same as that of Na$_2$Ni$_2$TeO$_6$. 
Simply, $T_\mathrm{N}$ decreases. 
No dimensional crossover is observed.  

\subsection{Below $T_\mathrm{N}$}

\begin{figure}[t]
 \begin{center}
 \includegraphics[width=0.80\linewidth]{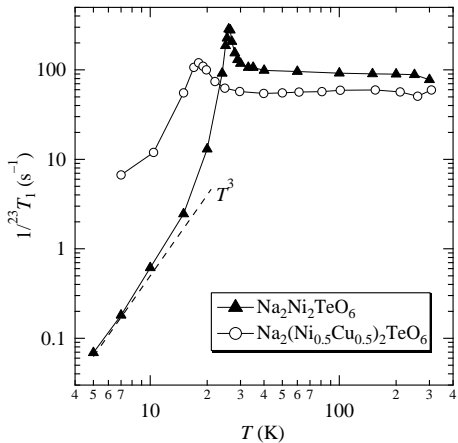}
 \end{center}
 \caption{\label{f6}
Log-log plots of 1/$^{23}T_1$ against temperature for Na$_2$Ni$_2$TeO$_6$ and Na$_2$(Ni$_{0.5}$Cu$_{0.5}$)$_2$TeO$_6$. 
A broken line indicates a function of eq.~(\ref{BP}). Solid curves are visual guides. 
 }
 \end{figure}
 
Figure~\ref{f6} shows log-log plots of 1/$^{23}T_1$ against temperature for Na$_2$Ni$_2$TeO$_6$ and Na$_2$(Ni$_{0.5}$Cu$_{0.5}$)$_2$TeO$_6$. 
With cooling down below $T_\mathrm{N}$, 1/$^{23}T_1$ rapidly decreases.  
The broken line indicates a $T^3$ function as a visual guide.  
In conventional antiferromagnetic states, the nuclear spin transitions are caused by Raman scattering and three magnon scattering~\cite{BP}.  
Then, 1/$T_1$ is expressed by 
\begin{eqnarray}   
{1\over {T_1}} \propto \Bigl({T\over {T_\mathrm{N}}}\Bigr)^3 
\label{BP}
\end{eqnarray}  
in a temperature range of $T_\mathrm{N} > T \gg T_{AE}$, where $T_{AE}$ corresponds to an energy gap in the spin wave spectrum~\cite{BP}.
The energy gap is due to a crystalline anisotropy field. 
The rapid drop of 1/$^{23}T_1$ below $T_\mathrm{N}$ results from the suppression of low energy excitations by the energy gap. 
Below $T_{AE}$, an activation-type temperature dependence
should be observed in 1/$T_1$. 
Since no activation behavior was observed down to 5 K, one may estimate $T_{AE} <$ 5 K.      

\section{Conclusions}

In conclusion, we found three dimensional critical phenomenon near $T_\mathrm{N}$ = 26 K for Na$_2$Ni$_2$TeO$_6$ and $T_\mathrm{N}$ = 18 K for Na$_2$(Ni$_{0.5}$Cu$_{0.5}$)$_2$TeO$_6$ from measurements of the $^{23}$Na nuclear spin-lattice relaxation rate 1/$^{23}T_1$. 
We have analyzed the NMR results by Ni$^{2+}$ with $S$ = 1 and obtained sound values of parameters for Na$_2$Ni$_2$TeO$_6$. 
We attribute the deviation from the Curie-Weiss law and the broad maximum of uniform magnetic susceptibility to two dimensional spin-spin correlation on a honeycomb lattice.  

The author thanks M. Isobe (Max Planck Institute) for X-ray diffraction measurements, K. Morimoto, C. Michioka, K. Yoshimura (Kyoto University) for sample preparation and characterization at an early stage.

\end{document}